\documentclass[12pt]{article}

\usepackage{subfig} 
\usepackage{url}
\usepackage{wrapfig}
\usepackage{eufrak} 
\usepackage{amsmath}   
\usepackage{a4p}
\usepackage{microtype}\usepackage{lmodern}\usepackage[T1]{fontenc}
\usepackage{mathptmx}
\usepackage{graphicx}
\usepackage[normalem]{ulem} 
\usepackage{color}
\definecolor{myred}{rgb}{0.6,0,0} 
\definecolor{myblue}{rgb}{0,0.2,0.4}
\definecolor{mygreen}{rgb}{0,0.9,0.1}
\definecolor{Orange}{rgb}{1.,0.65,0.}
\usepackage{color}
\definecolor{myred}{rgb}{1.0,0,0} 
\definecolor{mygreen}{rgb}{0,0.9,0.1} 
\definecolor{myblue}{rgb}{0,0.2,0.4}
\definecolor{mygray}{rgb}{.8,.8,.8}
\definecolor{darkorange}{rgb}{1, 0.549, 0}
\definecolor{purple}{rgb}{0.6,0.4,0.6}
\definecolor{mymagenta}{rgb}{0.6,0.4,0.6}
 \definecolor{LightCyan} {rgb}{0.88,1.,1.}
 \definecolor{Orange} {rgb}{1.,0.65,0.}
 \definecolor{PaleGreen} {rgb}{0.6,0.98,0.6}
 \definecolor{Pink} {rgb}{1.,0.75,0.8}
\definecolor{Red}{rgb}{1,0,0}
   \definecolor{Blue}{rgb}{0,0,1}
   \definecolor{Yellow}{rgb}{1,1,0}
   \definecolor{Orange}{rgb}{1,0.4,0}
   \definecolor{Pink}{rgb}{1,0,1}
   \definecolor{Purple}{rgb}{0.5,0,0.5}
   \definecolor{Teal}{rgb}{0,0.5,0.5}
   \definecolor{Navy}{rgb}{0,0,0.5}
   \definecolor{Aqua}{rgb}{0,1,1}
   \definecolor{Lime}{rgb}{0,1,0}
   \definecolor{Green}{rgb}{0,0.5,0}
   \definecolor{Olive}{rgb}{0.5,0.5,0}
   \definecolor{Maroon}{rgb}{0.5,0,0}
   \definecolor{Brown}{rgb}{0.6,0.4,0.2}
   \definecolor{Black}{gray}{0}
   \definecolor{Gray}{gray}{0.5}
   \definecolor{Silver}{gray}{0.75}
   \definecolor{White}{gray}{1}

\usepackage{hyperref}  
\definecolor{darkblue}{rgb}{0,0,.5}
\hypersetup{
linktocpage=true,
    bookmarks=true,         
breaklinks=true,
    frenchlinks=false, %
    unicode=false,          
    pdftoolbar=true,        
    pdfmenubar=true,        
    pdffitwindow=false,     
    pdfstartview={FitH},    
    pdfauthor={J.Fleischer, T.Riemann},     
    pdfcreator={JF+TR},   
    pdfproducer={JF+TR}, 
    pdfkeywords={}, 
    pdfnewwindow=true,      
    colorlinks=false,       
    linkcolor=red,          
    citecolor=green,        
    filecolor=magenta,      
    urlcolor=cyan           
}
\usepackage[all]{hypcap} 

\usepackage{rotating}

\numberwithin{equation}{section}
\numberwithin{figure}{section}
\numberwithin{table}{section}

\newcommand{\be}{\begin{equation}}
\newcommand{\ee}{\end{equation}}
\newcommand{\bea}{\begin{eqnarray}}
\newcommand{\eea}{\end{eqnarray}}

\newcommand{\nl}{\nonumber \\}

\def \eps {\varepsilon}

\newcommand{\nn}{\nonumber}

\begin{document}
\sloppy

\texttt{
\begin{flushleft}
DESY 11-063 
\\
BI-TP 2011/12
\\
SFB/CPP-11-25
\\
LPN 11-26
\end{flushleft}
}
\vspace{0.3cm}

\bigskip

\begin{center}
{\LARGE \bf
Calculating contracted tensor Feynman integrals}
\\
\vspace{1.0cm}

\renewcommand{\thefootnote}{\fnsymbol{footnote}}
{\Large J. Fleischer~\footnote{E-mail:~fleischer@physik.uni-bielefeld.de}${}^{a}$  ~~and~~
        T. Riemann~\footnote{E-mail:~Tord.Riemann@desy.de}${}^{b}$        }
\\[1cm]
\end{center}

{\noindent
{${}^{a}$~Fakult\"at f\"ur Physik, Universit\"at Bielefeld, Universit\"atsstr. 25,  33615
Bielefeld, Germany }
\\\noindent
{${}^{b}$~Deutsches Elektronen-Synchrotron, DESY, Platanenallee
  6, 15738 Zeuthen, Germany}
}

\vspace{1cm}

\begin{center}
{\Large \bf
{Abstract}}
\\[5mm]\end{center}
A recently derived approach to the tensor reduction  of 5-point one-loop Feynman integrals expresses the tensor coefficients by scalar 1-point to 4-point Feynman integrals completely algebraically.
In this letter we derive extremely compact algebraic expressions for the contractions of the tensor integrals with external momenta.
This is based on sums over signed minors weighted with scalar products of the external momenta.
With these contractions one can construct the invariant amplitudes of the matrix elements under consideration,
and the evaluation of one-loop contributions to massless  and massive multi-particle production at high energy colliders like LHC and ILC is expected to be performed very efficiently.

\bigskip

PACS index categories: 12.15.Ji, 12.20.Ds, 12.38.Bx

\setcounter{footnote}{0}
\renewcommand{\thefootnote}{\arabic{footnote}}

\section{\label{Intro} Introduction}
In a recent article  \cite{Fleischer:2010sq} (hereafter quoted as reference I), we have worked out an algebraic method to present one-loop tensor integrals in terms of  scalar one-loop $1$-point to 
$4$-point functions.
The tensor integrals are defined as
\bea
\label{definition}
 I_{n}^{\mu_1\cdots\mu_R} &=&~\int \frac{d^dk}{i {\pi}^{d/2}}~~\frac{\prod_{r=1}^{R} k^{\mu_r}}{\prod_{j=1}^{n}c_j},
\eea
with denominators $c_j$, having  \emph{chords}
$q_j$,
\begin{eqnarray}\label{propagators}
c_j &=& (k-q_j)^2-m_j^2 +i \epsilon.
\end{eqnarray}
Here, we use the generic dimension $d=4-2\epsilon$.
The central problem are the 5-point tensor functions.
We derived algebraic expressions for them in terms of \emph{higher-dimensional} scalar 4-point functions with \emph{raised indices} (powers of the scalar propagators).
There are several ways to go.
One option is to avoid the appearance of inverse Gram determinants $1/()_5$. 
For rank $R=5$, e.g.,
\bea\label{compl5a}
I_{5}^{\mu\, \nu\, \lambda \rho \sigma} &=&  
\sum_{s=1}^5 \Biggl[
\sum_{i,j,k,l,m=1}^{5}  q_i^{\mu} q_j^{\nu}  q_k^{\lambda} q_l^{\rho} q_m^{\sigma} E_{ijklm}^s
+
\sum_{i,j,k=1}^5 g^{[\mu \nu} q_i^{\lambda} q_j^{\rho} q_k^{\sigma]} E_{00ijk}^s
+
\sum_{i=1}^5
g^{[\mu \nu} g^{\lambda \rho} q_i^{\sigma]} E_{0000i}^s \Biggr],
\nl
\label{compl5}
\eea
see equations (I.4.60), (I.4.61).
The tensor coefficients 
are expressed in terms of integrals $I_{4,i\cdots}^{[d+]^l,s}$, e.g. according to (I.4.62):
\bea
E_{ijklm}^s &=&
-\frac{1}{{0\choose 0}_5}  \left\{ \left[
{0l\choose sm}_5 {n}_{ijk}I_{4,ijk}^{[d+]^4,s}+(i \leftrightarrow l)+(j \leftrightarrow l)+
(k \leftrightarrow l)\right]
+{0s\choose 0m}_5{n}_{ijkl} I_{4,ijkl}^{[d+]^4,s} \right\} .
\nl
\label{Ewxyz5}
\eea
The scalar integrals are
\begin{eqnarray}
  \label{eq:Inij}
   I_{p, \, i\,j \,k\cdots} ^{[d+]^l,stu \cdots} =  \int \frac{d^{[d+]^l}k}{i\pi^{[d+]^l/2}}  \prod_{r=1}^{n} \, \frac{1}{c_r^{1+\delta_{ri} + \delta_{rj}+\delta_{rk}+\cdots
                -\delta_{rs} - \delta_{rt}-\delta_{ru}-\cdots}} ,
\end{eqnarray}
where $p$ is the number of internal lines and $[d+]^l=4-2 \eps+2 l $.
Further, we use the notations of signed minors (I.2.14).
At this stage, the higher-dimensional 4-point integrals still depend on tensor indices, namely through the indices $i,j$ etc.
The most complicated explicit example $I_{4,ijkl}^{[d+]^4,s}$ appears in \eqref{Ewxyz5}.
Now, in a next step, one may avoid the appearance of inverse sub-Gram determinants $()_4$.
Indeed, after tedious manipulations, one arrives at representations in terms of scalar integrals $I_4^{[d+]^l}$ plus simpler 3-point and 2-point functions,
and the complete dependence on the indices $i$   of the tensor coefficients is contained now in the pre-factors with signed minors. One can say that the indices \emph{decouple} from the integrals. 
As an example, we reproduce the 4-point part of (I.5.21),
\bea
&&{n}_{ijkl} I_{4,ijkl}^{[d+]^4}=
\frac{{0\choose i}}{{0\choose 0}}
\frac{{0\choose j}}{{0\choose 0}}\frac{{0\choose k}}{{0\choose 0}}\frac{{0\choose l}}{{0\choose 0}}
d(d+1)(d+2)(d+3)I_4^{[d+]^4} \nn \\
&&+~\frac{{0i\choose 0j}{0\choose k}{0\choose l}+{0i\choose 0k}{0\choose j}{0\choose l}+{0j\choose 0k}{0\choose i}{0\choose l}+
{0i\choose 0l}{0\choose j}{0\choose k}+{0j\choose 0l}{0\choose i}{0\choose k}+{0k\choose 0l}{0\choose i}{0\choose j}}
{{0\choose 0}^3}  d(d+1)I_4^{[d+]^3}\nn \\
&&+~\frac{{0i\choose 0l}{0j\choose 0k}+{0j\choose 0l}{0i\choose 0k}+{0k\choose 0l}{0i\choose 0j}}
{{0\choose 0}^2}I_4^{[d+]^2} 
~~+~ \cdots
\label{fulld4}
\eea
In \eqref{fulld4}, one has to understand the 4-point integrals to carry the corresponding index $s$ and the signed minors are ${0\choose k}\to {0s\choose ks}_5$ etc. This type of relations may be called
``recursion relations for small Gram determinants''.

\bigskip

In an alternative treatment, tensor reduction formulas free of $g^{\mu\nu}$ terms were derived in ref. I.
In that case, inverse powers of $()_5$ are tolerated.
The most involved object studied was (I.3.20):
\bea\label{3.52}
I_5^{\mu \nu \lambda\rho\sigma}=
I_5^{\mu\nu\lambda\rho} \cdot Q_0^{\sigma}
-\sum_{s=1}^{5} I_4^{\mu\nu\lambda\rho, s} \cdot Q_s^{\sigma},
\eea
with the 4-point tensor functions (I.3.29) and (I.3.18)
\bea\label{3.363}
I_4^{\mu \nu \lambda \rho, s}&=&
Q_0^{s,\mu} Q_0^{s,\nu}Q_0^{s,\lambda}Q_0^{s,\rho}  I_4^s+
3 \frac{{\left( \right)}_5^2}{{s\choose s}_5^2} Q_s^{\mu}Q_s^{\nu}Q_s^{\lambda}Q_s^{\rho}
\cdot I_4^{[d+]^2,s}
+\frac{{\left( \right)}_5}{{s\choose s}_5} Q_{s}^{[\mu}Q_{s}^{\nu}J_4^{\lambda,s } Q_0^{s,\rho ]} 
~~+~\cdots ,
\nl
\\
J_4^{\mu, s}&=&
-Q_0^{s,\mu} I_4^{[d+]} +
\sum_{t=1}^{5} Q_t^{s,\mu} I_3^{[d+],st} .
\label{J4V}
\eea
The dots in \eqref{3.363} indicate 3-point functions, and 
\bea
 Q_s^{\mu}&=&\sum_{i=1}^{5}  q_i^{\mu} \frac{{s\choose i}_5}{\left(  \right)_5},~~~ s=0 \cdots 5,
\label{Q6}
\\\label{3.11}
Q_s^{t,\mu}&=&\sum_{i=1}^{5} q_i^{\mu} \frac{ {st\choose it}_5}{{t\choose t}_5},~~~ s,t=1 \cdots 5.
\eea
Also here, the tensor coefficients have been represented by scalar functions free of tensor indices.

We remark that all the above-mentioned results are due to a systematic application
of methods described and developed in \cite{Fleischer:1999hq}. For the present paper
the ``algebra of signed minors'' \cite{Melrose:1965kb} plays a particularly important role.
This method was also used in \cite{Fleischer:1999hq} and further developed to
its full power in \cite{Diakonidis:2008ij}. In the latter article also 6-point functions
have been treated on this basis.

In the next section we will develop a very efficient method to evaluate realistic matrix elements with tensor integral representations of the above kind. 

\section{\label{5to4}Contracting the tensor integrals}
To apply the approach most efficiently one should construct projection operators
for
the invariant amplitudes of the matrix elements under consideration. These projectors,
of course, depend on the tensor basis and have to be constructed for each
process specifically. 
If done like that, the tensor indices of the loop integrals 
are saturated by contractions with external momenta $p_r$. 
The chords in \eqref{propagators} are given in terms
of the external momenta as $q_i=-(p_1+p_2+\cdots +p_i)$, with $q_n=0$, and inversely 
$p_r = q_{r-1}-q_r $. Then,  any of the integrals to be evaluated is a simple linear combination of integrals containing products with chords, $(q_r \cdot k)$:
\bea
\label{definition2}
q_{i_1\mu_1}\cdots q_{i_R\mu_R} ~~I_{5}^{\mu_1\cdots\mu_R} 
&=&
~\int\frac{ d^d k~}{i {\pi}^{d/2}}~\frac{\prod_{r=1}^{R} (q_{i_r} \cdot k)}{\prod_{j=1}^{5}c_j}.
\eea
There is another type of external vectors, i.e. the polarisation vectors $\epsilon_i$ of spin-$1$ bosons. 
They, however, are taken into account in the definition 
of the tensor structure of the matrix elements in terms of scalar
products $({\epsilon}_i \cdot p_j)$ with some external momenta $p_j$.
The same applies to contractions with $\gamma$ matrices 
~${\epsilon}_i  \hspace*{-3mm} /$~ and ~$p_i  \hspace*{-3mm} /$ ~in spinor chains.
Thus, polarisation vectors and $\gamma$ matrices will not show up in the sums one has to perform. 

If the integration momentum $k$ is self-contracted, one may use the identity
$k^2/c_j= 1 + m_j^2/c_j + 2(q_j\cdot k)/c_j$
in order to transform the integral to the type \eqref{definition2} plus simpler ones. Since the approach
uses $q_n=0$, one should take care that $c_5$ is not canceled. Then
the procedure can also be applied to the scratched 4-point functions in the same manner
as for the 5-point functions. 
Nevertheless, we do not consider this  approach
as optimal since in any case many new terms are produced and it seems more adequate
also in this case to perform the corresponding sums as indicated in appendix A
(see.\ref{eq-wa-83x} - \ref{eq-wa-83zfin}).

\bigskip

We will represent now the integrals  \eqref{definition2} as compact linear combinations of scalar one-loop integrals, in higher dimensions and with indices 1.
In order to demonstrate the method, we will explicitly work out only the simplest cases with rank $R \leq 3$, but we will collect all the sums over signed minors needed also for the most complicated cases exemplified in the introduction.

\bigskip

The tensor $5$-point integral of rank $R=1$  (I.4.6) yields, when contracted with a chord,
\begin{eqnarray}\label{i5vc1}
q_{a \mu} I_{5}^{\mu}&=& - \frac{1 }{{0\choose 0}_5} 
\sum_{s=1}^{5}
\left[
\sum_{i=1}^{4} (q_{a} \cdot q_i) {0i\choose 0s}_5 \right]
 I_{4}^{s}.
\label{i5vc2}
\end{eqnarray}
In fact, the sum over $i$ may be performed explicitly, it is the sum ${\Sigma}_{a}^{1,s}$ \eqref{eq-wa-83a} 
 listed in appendix \ref{Simplify}, and we get immediately
\bea
q_{a \mu}  I_{5}^{\mu}= 
-~\frac{1 }{{0\choose 0}_5}\sum_{s=1}^{5}{\Sigma}_{a}^{1,s}~~I_{4}^{s}.
\eea

\bigskip

The tensor $5$-point integral of rank $R=2$ (I.4.19),
\begin{eqnarray}
I_{5}^{\mu\, \nu\,}= \sum_{i,j=1}^{4} \, q_i^{\mu}\, q_j^{\nu} E_{ij} +
g^{\mu \nu}  E_{00} ,
\label{final2}
\end{eqnarray}
has the following tensor coefficients free of $1/()_5$:
\bea\label{E00}
E_{00}
&=&
 - \sum_{s=1}^5    \frac{1}{2} \frac{1}{{0\choose 0}_5} {s\choose 0}_5 I_4^{[d+],s},
\\
E_{ij} 
\label{Exy}
&=&
\sum_{s=1}^5   \frac{1}{{0\choose 0}_5}   \left[{0i\choose sj}_5 I_4^{[d+],s}+
{0s\choose 0j}_5 I_{4,i}^{[d+],s} \right].
\label{Eij}
\eea
Equation \eqref{final2}  yields for the contractions with chords:
\bea\label{qqi2}
q_{a \mu} q_{b \nu} I_{5}^{\mu\, \nu\,}= \sum_{i,j=1}^4 (q_a \cdot q_i) (q_b \cdot q_j) E_{ij}+
(q_a \cdot q_b) E_{00} .
\eea
Applying (\ref{eq-wa-82}) on the first part of (\ref{Eij}), it is easy to see that the
term with $(q_a \cdot q_b)$  cancels the $E_{00}$. 
In a next step the $I_{4,i}^{[d+],s}$ may be eliminated by a scratched version of (I.A.6) or of (I.5.10). 
We use here the latter one which is free of $1/()_4$:
\bea
I_{4,i}^{[d+]^l,s} 
&=&
\frac{1}{{0s\choose 0s}_5}\left[-{0s\choose is}_5  (d+2l-5) I_4^{[d+]^l,s}
+\sum_{t=1}^5 {0st\choose 0si}_5 I_3^{[d+]^{l-1},st}\right],~~l=1,2.
\label{I4id+2a}
\eea
We again arrived at a  representation where sums over $i,j$ are decoupled from the scalar master integrals.
Equations  \eqref{eq-wa-83a} - \eqref{eq-wa-82} may be applied,
and the contribution of $I_4^{[d+],s}$ to \eqref{qqi2} reads now
\bea
\frac{1}{{0\choose 0}_5} \left\{ \left\{{\Sigma}_{ab}^{1,s}\right\}_{sp}-
\frac{1}{{0s\choose 0s}_5} {\Sigma}_{b}^{1,s} \cdot {\Sigma}_{a}^{2,s}\right\} I_4^{[d+],s},
\eea
where $\left\{{\Sigma}_{ab}^{1,s}\right\}_{sp}$ is the $(q_a \cdot q_b)$-independent part of ${\Sigma}_{ab}^{1,s}$, \eqref{eq-wa-82}.
The ${\Sigma}_{b}^{1,s}$ and ${\Sigma}_{a}^{2,s}$ are given in (\ref{eq-wa-83a}) and (\ref{eq-wa-83b}),
respectively. 
A further simplification can be achieved with the identity
\bea
{\left(\right)}_5+\frac{{s\choose 0}_5^2}{{0s\choose 0s}_5}=\frac{{0\choose 0}_5 {s\choose s}_5}
{{0s\choose 0s}_5} ,
\eea
 and finally \eqref{qqi2} simply reads
\bea
q_{a \mu} q_{b \nu} I_{5}^{\mu\, \nu\,}
&=&\frac{1}{4}
\sum_{s=1}^{5}\Biggl\{
\frac{{s\choose 0}_5}{{0s\choose 0s}_5}({\delta}_{ab}{\delta}_{as}+{\delta}_{5s}) 
+\frac{{s\choose s}_5}{{0s\choose 0s}_5} \Bigl[ 
\left({\delta}_{as}-{\delta}_{5s}\right) \left(Y_{b5}-Y_{55} \right)
\nl &&
+ ~
\left({\delta}_{bs}-{\delta}_{5s}\right) \left(Y_{a5}-Y_{55} \right)+
\frac{{s\choose 0}_5}{{0\choose 0}_5}\left(Y_{a5}-Y_{55} \right)\left(Y_{b5}-Y_{55} \right) \Bigr]
\Biggr\}I_4^{[d+],s} 
\nl &&
+~\frac{1}{{0\choose 0}_5} \sum_{s=1}^{5} \frac{{\Sigma}^{1,s}_{b}}{{0s\choose 0s}_5} \sum_{t=1}^{5}
{\Sigma}^{2,st}_{a} I_3^{st} ,
\label{contr2}
\eea
with ${\Sigma}^{2,st}_{a}$ given in (\ref{eq-wa-84b}).

The result is typical in the sense that, after  summation over the tensor indices,
terms with factors ${s\choose s}_5$ will appear, i.e.  with the Gram determinants $()_4$ of the $4$-point functions.
This circumstance is advantageous when reducing the dimensional integrals $I_4^{[d+]^l,s}$ to lower dimensions, where factors $1/()_4$ are produced.
So, the problem of the small $4$-point Gram determinants, {discussed in great detail in ref.~I},  is at least partially eliminated. 
The remaining
terms are factored by Kronecker's $\delta$-symbol and yield contributions for specific indices $a,b$ only -- after summation over $s$. 

\bigskip

Finally, we exemplify the rank $R=3$ case.
The tensor can be written as follows (see (I.4.35)-(I.4.37)):
\begin{eqnarray}
I_{5}^{\mu\, \nu\, \lambda}&&= \sum_{i,j,k=1}^{5} \, q_i^{\mu}\, q_j^{\nu} \, q_k^{\lambda}
E_{ijk}+\sum_{k=1}^5 g^{[\mu \nu} q_k^{\lambda]} E_{00k},
\label{Exyz0}
\end{eqnarray}
with
\bea
\label{Exyz1}
E_{00k} 
&=& \sum_{s=1}^5 \frac{1}{{0\choose 0}_5}  \left[\frac{1}{2} {0s\choose 0k}_5 I_4^{[d+],s}- \frac{d-1}{3} {s\choose k}_5 I_4^{[d+]^2,s} \right]  ,
\\
E_{ijk} 
&=&-  \sum_{s=1}^5\frac{1}{{0\choose 0}_5}  \left\{ \left[{0j\choose sk}_5 I_{4,i}^{[d+]^2,s}+
(i \leftrightarrow j)\right]+{0s\choose 0k}_5 {\nu}_{ij} I_{4,ij}^{[d+]^2,s} \right\}.
\label{Exyz2}
\eea
We eliminate now indices from scalar integrals with recursion \eqref{I4id+2a} and further recursion relations 
applicable for cases with small Gram determinants $()_4$, reproduced here in the unscratched 
forms~(I.5.15) and~(I.5.16):
\bea
{\nu}_{ij} I_{4,ij}^{[d+]^2}&=&  
\frac{{0\choose i}}{{0\choose 0}}\frac{{0\choose j}}{{0\choose 0}}
(d-2)(d-1)I_4^{[d+]^2}+\frac{{0i\choose 0j}}{{0\choose 0}}I_{4}^{[d+]}\nl
&&-~\frac{{0\choose j}}{{0\choose 0}}\frac{d-2}{{0\choose 0}}\sum_{t=1}^4 {0t\choose 0i}I_3^{[d+],t} ~~~+
\frac{1}{{0\choose 0}}\sum_{t=1}^4 {0t\choose 0j}I_{3,i}^{[d+],t} ,
\label{want1}
\\
I_{3,i}^{[d+],t}&=&
-\frac{{0t\choose it}}{{0t\choose 0t}}(d-2) I_{3}^{[d+],t}+\frac{1}{{0t\choose 0t}}
\sum_{u=1}^4 {0tu\choose 0ti}I_2^{tu} .
\label{indices}
\eea
The ${0t\choose 0t}$ in \eqref{indices} vanishes for infrared divergent $3$-point functions
and therefore 
one has to use ``standard'' recursions a la (I.A.10) in this case. 
Anyway, such problems are not the
concern of this letter and they have to be discussed separately if met.

After these preparations
we can now evaluate the contractions of the tensor with three chords:
\bea
\label{Proj3}
q_{a \mu} q_{b \nu} q_{c \lambda}I_{5}^{\mu\, \nu\, \lambda}= 
&&\sum_{i,j,k=1}^4 
(q_a \cdot q_i) (q_b \cdot q_j) (q_c \cdot q_k)E_{ijk} \\
&&+~ \sum_{k=1}^4 \left[(q_a \cdot q_b) (q_c \cdot q_k) +(q_a \cdot q_c) (q_b \cdot q_k)+(q_b \cdot q_c) (q_a \cdot q_k)\right] E_{00k} . \nn
\eea
For the triple sum over $i,j,k$ in \eqref{Proj3} we get
\bea
\label{tensor3}
-\frac{1}{{0\choose 0}_5}\sum_{s=1}^5
\Bigl\{
{\Sigma}^{1,s}_{bc}\cdot \sum_{i=1}^{4} (q_a \cdot q_i) I_{4,i}^{[d+]^2,s}
+
(b \leftrightarrow a)
+ 
{\Sigma}^{1,s}_{c}\cdot \sum_{i,j=1}^{4} (q_a \cdot q_i) (q_b \cdot q_j) {\nu}_{ij} I_{4,ij}^{[d+]^2,s} \Bigr\},
\eea
and get further for the sums in \eqref{tensor3}
\bea
\sum_{i=1}^{4} (q_a \cdot q_i) I_{4,i}^{[d+]^2,s} &=&
\frac{1}{{0s\choose 0s}_5}\left\{-{\Sigma}^{2,s}_{a}(d-1)I_{4}^{[d+]^2,s}+\sum_{t=1}^{5} {\Sigma}^{2,st}_{a}I_3^{[d+],st}
\right\},  
\\ 
\sum_{i,j=1}^{4} (q_a \cdot q_i) (q_b \cdot q_j) {\nu}_{ij} I_{4,ij}^{[d+]^2,s}
&=&
\frac{1}{{0s\choose 0s}^2_5}{\Sigma}^{2,s}_{a}{\Sigma}^{2,s}_{b}(d-2)(d-1)I_{4}^{[d+]^2,s}+
\frac{1}{{0s\choose 0s}_5}{\Sigma}^{3,s}_{ab}I_{4}^{[d+],s}  
\nonumber\\
&&
-~\frac{1}{{0s\choose 0s}_5}\sum_{t=1}^{5}\left\{\frac{1}{{0s\choose 0s}_5}{\Sigma}^{2,s}_{b}{\Sigma}^{2,st}_{a}(d-2)I_3^{[d+],st}
\right. 
 \\ &&
+\left.~
\frac{1}{{0st\choose 0st}_5}{\Sigma}^{2,st}_{b} 
\left[{\Sigma}^{3,st}_{a}(d-2)I_3^{[d+],st}-\sum_{u=1}^{5}{\Sigma}^{2,stu}_{a}I_2^{stu} \right] \right\}.\nn
\eea
Finally, for the single sum in \eqref{Proj3} we have
\bea
\sum_{k=1}^4 (q_c \cdot q_k) E_{00k}^s=\frac{1}{2{0\choose 0}_5} \left[{\Sigma}^{1,s}_{c}I_{4}^{[d+],s}
-\frac{d-1}{3} \left( \right)_5 \left({\delta}_{cs}-{\delta}_{5s}\right)I_{4}^{[d+]^2,s} \right].
\eea
We leave the task to collect the terms needed in \eqref{Proj3} to the reader. We only mention
that similar simplifications like those for the tensor of rank $R=2$ can be achieved if one
evaluates \eqref{Proj3} with a symmetrized version of \eqref{Exyz2}.

It is interesting to compare our approach with the so called OPP method \cite{delAguila:2004nf,Ossola:2006us}. For this purpose we
concentrate on the $5$-point function, which was discussed in detail so far.
Both methods start from a recursion relation, namely (2.2) in \cite{delAguila:2004nf}
and (I.2.5), derived in \cite{Diakonidis:2009fx}. In further steps, of course, we do 
not identify the results, but find analogies. 

The first analogy is the representation of the $5$-point tensor
by means of the number of scalar propagators, resulting in $4$-, $3$-, $2$- and $1$-point
functions. This is given in \cite{Ossola:2006us} by (1.1) and (1.2). In the present work
we {use representations where the} tensors are reduced correspondingly to $4$-, $3$-, $2$- and 
$1$-point integrals (with indices 1), however the integrations are performed in general
in higher {space-time} dimensions. In fact there will occur, in general, even several integrals in different
dimension, like e.g. in  \eqref{fulld4}.
One essential difference is that in our approach there is 
no $0$-point, ``spurious'' contribution: performing recursions, these
terminate with $1$-point functions;
see e.g. appendix~A of ref.~I.

The next step is the analogy of the coefficients $d,c,b,a$ in \cite{Ossola:2006us}
and ours, given in ref.~I. Taking again \eqref{fulld4} as an example, {our coefficients are written}
explicitly in terms of \emph{signed minors} - as can be seen from ref.~I for the $3$-, $2$- and
$1$-point functions as well. This means we do not need a recursion going down the 
chain $d,c,b,a$. Instead, we have \emph{solved} the recursion. In \cite{Ossola:2006us} the tensor
indices are carried by massless $4$-vectors $\textit{l}_1 \dots \textit{l}_4$ while in our case
they are carried by the chords $q_i$.

In \cite{Ossola:2006us} the coefficients $d,c,b,a$ are calculated numerically, {while} here
they are given analytically.
{So we can go one step further and perform} the
summation over the indices as demonstrated above and in detail in appendix A. 
In fact,
{relying on projectors} to obtain the invariant amplitudes of a matrix element, these
sums are at most two-fold.
The reason that there are no further sums to be evaluated
is due to the fact that not only the indices \emph{decouple} from the integrals, but
{in addtion to} that they also  \emph{factorize} such that at most two indices occur in any one signed minor.

\section{\label{conclude}Conclusions}

The contracted rank $R=1\cdots 3$ tensors of the 5-point function have been expressed by scalar integrals, accompanied by compact expressions for sums over products of chords and signed minors.
It is evident how the general case has to be treated, once a table of sums as given in the appendix is available. 
The scalar integrals may be defined in higher dimensions or in the generic dimension, depending on the preferred algorithm of the final numerical evaluations and on questions related to a treatment (or avoidance) of inverse Gram determinants.

Based on the approach defined in this letter, we expect a considerable economization of 
cross-section calculations in cases where an essential part of the computational time
and storage is spent on tensor reduction.

\section*{Acknowledgements}
We acknowledge useful discussions with S. Moch and P. Uwer.
J.F. thanks DESY for kind hospitality.
Work is supported in part by Sonderforschungsbereich/Trans\-re\-gio SFB/TRR 9 of DFG
``Com\-pu\-ter\-ge\-st\"utz\-te Theoretische Teil\-chen\-phy\-sik" and European Initial Training Network LHCPHENOnet PITN-GA-2010-264564.

\appendix
\allowdisplaybreaks
\section{Sums over contracted chords and signed minors}\label{Simplify}
A useful notation is
\bea
\label{gram}
Y_{ij}=-(q_i-q_j)^2+m_i^2+m_j^2.
\eea
The simplest contractions are given in \cite{Diakonidis:2009fx}:
\bea
(q_i \cdot Q_0) &=&\sum_{j=1}^{n-1} (q_i  \cdot q_j) \frac{{0\choose j}_n}{{\left(\right)}_n}=-\frac{1}{2}\left(
 Y_{in}-Y_{nn} \right), ~~~i=1, \dots , n-1,
\label{Scalar1}
\\\nl
(q_i \cdot Q_s) &=&\sum_{j=1}^{n-1} (q_i  \cdot q_j) \frac{{s\choose j}_n}{{\left( \right)}_n}=\frac{1}{2}\left(
{\delta}_{is}-{\delta}_{ns}\right), ~~~i=1, \dots , n-1, ~~~s=1, \dots n.
\label{Scalar2}
\eea
The $Q_s,  Q_0$ are defined in \eqref{Q6}.
In~(\ref{Scalar1}) and~(\ref{Scalar2}), $q_n=0$ is assumed  since only in this case the relation
\bea\label{Scalar2a}
(q_i \cdot q_j) =\frac{1}{2} \left[Y_{ij}-Y_{in}-Y_{nj}+Y_{nn} \right]
\eea
holds which is needed for their derivations.

Further sums are needed if the $4$-point tensors are contracted:
\bea 
\label{eq-wa-83a}
{\Sigma}^{1,s}_{a} 
&\equiv&
\sum_{i=1}^{4}(q_a \cdot q_i)  {0s\choose 0i}_5 
~=~ +\frac{1}{2} \left\{
{s\choose 0}_5\left(Y_{a5}-Y_{55} \right)+
{0\choose 0}_5\left({\delta}_{as}-{\delta}_{5s}\right) \right\}, 
\\
\label{eq-wa-83b}
{\Sigma}^{2,s}_{a} &\equiv&
\sum_{i=1}^{4}(q_a \cdot q_i)  {0s\choose is}_5 
~=~
-\frac{1}{2} \left\{
{s\choose s}_5\left(Y_{a5}-Y_{55}\right) +
{s\choose 0}_5\left({\delta}_{as}-{\delta}_{5s}\right)
\right\}.
\eea

Double sums for 4-point functions:
\bea\label{eq-wa-82}
{\Sigma}^{1,s}_{ab} &\equiv&
\sum_{i,j=1}^{4} (q_a \cdot q_i) (q_b \cdot q_j) {0i\choose sj}_5~=~\frac{1}{2}(q_a \cdot q_b){s\choose 0}_5
+\frac{1}{4}{\left(\right)}_5\left(Y_{b5}-Y_{55} \right)\left({\delta}_{as}-{\delta}_{5s}\right),
\\
\label{eq-wa-82x}
{\Sigma}^{2,s}_{ab} &\equiv&
\sum_{i,j=1}^{4} (q_a \cdot q_i) (q_b \cdot q_j) {si\choose sj}_5~=~\frac{1}{2}(q_a \cdot q_b){s\choose s}_5
-\frac{1}{4}{\left(\right)}_5\left({\delta}_{ab}{\delta}_{as}+{\delta}_{5s} \right),
\\
{\Sigma}^{3,s}_{ab} &\equiv&
\sum_{i,j=1}^{4} (q_a \cdot q_i) (q_b \cdot q_j) {0si\choose 0sj}_5
~=~\frac{1}{2}(q_a \cdot q_b) 
{0s\choose 0s}_5-\frac{1}{4}\left\{{s\choose s}_5  \left(Y_{a5}-Y_{55} \right) \left(Y_{b5}-Y_{55} \right) 
\right. \nn \\
&&\left. +
{s\choose 0}_5 \left[
\left({\delta}_{as}-{\delta}_{5s}\right) \left(Y_{b5}-Y_{55} \right) +
\left({\delta}_{bs}-{\delta}_{5s}\right) \left(Y_{a5}-Y_{55} \right)\right]
+{0\choose 0}_5 \left({\delta}_{ab}{\delta}_{as}+{\delta}_{5s}\right)  
\right\}. \nn \\
\label{eq-wa-82a}
\eea

Sums for $3$-point functions: 
\bea\label{eq-wa-84a}
{\Sigma}^{1,st}_{a} &\equiv&
\sum_{i=1}^{4}(q_a \cdot q_i)  {ts\choose is}_5
~=~\frac{1}{2} \left(1-{\delta}_{st}\right)\left\{{s\choose s}_5\left({\delta}_{at}-{\delta}_{5t}\right)
-{s\choose t}_5\left({\delta}_{as}-{\delta}_{5s}\right)\right\} ,
\\
\label{eq-wa-84b}
{\Sigma}^{2,st}_{a} &\equiv&
\sum_{i=1}^{4}(q_a \cdot q_i)  {0st\choose 0si}_5
\nl
&=&\frac{1}{2} \left(1-{\delta}_{st}\right)\left\{{ts\choose 0s}_5\left(Y_{a5}-Y_{55} \right) +
{0s\choose 0s}_5\left({\delta}_{at}-{\delta}_{5t}\right)-
{0s\choose 0t}_5 \left({\delta}_{as}-{\delta}_{5s} \right) \right\},
\\
\label{eq-wa-84d}
{\Sigma}^{3,st}_{a} &\equiv&
\sum_{i=1}^{4}(q_a \cdot q_i)  {0st\choose ist}_5
\nl
&=&-\frac{1}{2} \left\{{st\choose st}_5 \left(Y_{a5}-Y_{55} \right)+{st\choose s0}_5
\left({\delta}_{at}-{\delta}_{5t}\right)+{st\choose 0t}_5 \left({\delta}_{as}-{\delta}_{5s} \right)
\right\}.
\eea
\newpage
Double sums for 3-point functions:
\bea\label{eq-wa-91}
{\Sigma}^{4,st}_{ab} &\equiv&
\sum_{i,j=1}^{4} (q_a \cdot q_i) (q_b \cdot q_j) {0sti\choose 0stj}_5
~=~ \frac{1}{2}(q_a \cdot q_b)
{0st\choose 0st}_5 
-\frac{1}{4}\left(1-{\delta}_{st}\right)
\nn \\
&&
\times~
\left\{{st\choose st}_5\left(Y_{a5}-Y_{55} \right)\left(Y_{b5}-Y_{55} \right)+
{0s\choose 0s}_5\left({\delta}_{ab}{\delta}_{at}+{\delta}_{5t}\right)
+{0t\choose 0t}_5\left({\delta}_{ab}{\delta}_{as}+{\delta}_{5s}\right) \right. 
\nn \\
&&\left. +~{st\choose s0}_5\left[\left({\delta}_{at}-{\delta}_{5t}\right)\left(Y_{b5}-Y_{55} \right)+
                                  \left({\delta}_{bt}-{\delta}_{5t}\right)\left(Y_{a5}-Y_{55} \right) \right]\right.
\nn \\
&&\left. +~{st\choose 0t}_5\left[\left({\delta}_{as}-{\delta}_{5s}\right)\left(Y_{b5}-Y_{55} \right)+
                                    \left({\delta}_{bs}-{\delta}_{5s}\right)\left(Y_{a5}-Y_{55} \right) \right] \right.
\nn \\
&&\left. -~{0s\choose 0t}_5
\left[\left({\delta}_{at}-{\delta}_{5t}\right)\left({\delta}_{bs}-{\delta}_{5s}\right)+
      \left({\delta}_{bt}-{\delta}_{5t}\right)\left({\delta}_{as}-{\delta}_{5s}\right)\right] \right\} ,
 \\
\label{eq-wa-92}
{\Sigma}^{4,stu}_{ab} &\equiv&
\sum_{i,j=1}^{4} (q_a \cdot q_i) (q_b \cdot q_j) {stui\choose stuj}_5
~=~
\frac{1}{2}(q_a \cdot q_b) {stu\choose stu}_5
-\frac{1}{4}\left(1-{\delta}_{st}\right)\left(1-{\delta}_{su}\right)\left(1-{\delta}_{tu}\right) \nn \\
&&\left\{~
 {st\choose st}_5\left({\delta}_{ab}{\delta}_{au}+{\delta}_{5u} \right)
+{su\choose su}_5\left({\delta}_{ab}{\delta}_{at}+{\delta}_{5t} \right)
+{tu\choose tu}_5\left({\delta}_{ab}{\delta}_{as}+{\delta}_{5s} \right) \right. \nn \\
&&\left. -~{st\choose su}_5
\left[\left({\delta}_{at}-{\delta}_{5t}\right)\left({\delta}_{bu}-{\delta}_{5u}\right)+
      \left({\delta}_{bt}-{\delta}_{5t}\right)\left({\delta}_{au}-{\delta}_{5u}\right)\right] \right.
\nn \\
&&\left. -~{ts\choose tu}_5\left[\left({\delta}_{au}-{\delta}_{5u}\right)\left({\delta}_{bs}-{\delta}_{5s} \right)+
\left({\delta}_{bu}-{\delta}_{5u}\right)\left({\delta}_{as}-{\delta}_{5s} \right) \right] \right.
\nn \\
&&\left. -~{us\choose ut}_5\left[\left({\delta}_{at}-{\delta}_{5t}\right)\left( {\delta}_{bs}-{\delta}_{5s}\right)+
\left({\delta}_{bt}-{\delta}_{5t}\right)\left({\delta}_{as}-{\delta}_{5s}\right) \right] \right\} .
\eea

Sums for $2$-point functions:
\bea
\label{eq-wa-86}
{\Sigma}^{1,stu}_{a} 
&\equiv&
\sum_{i=1}^{4}(q_a \cdot q_i)  {stu\choose sti}_5
~~=~
\frac{1}{2}\left(1- {\delta}_{su}\right)\left(1- {\delta}_{tu}\right) 
\nl
&&
\times~ \left\{
 {st\choose st}_5\left({\delta}_{au}-{\delta}_{5u} \right)~
-{st\choose su}_5\left({\delta}_{at}-{\delta}_{5t} \right)~
-{ts\choose tu}_5\left({\delta}_{as}-{\delta}_{5s} \right)  
\right\} ,
\\
\label{eq-wa-87}
{\Sigma}^{2,stu}_{a} 
&\equiv&
\sum_{i=1}^{4}(q_a \cdot q_i)  {0stu\choose 0sti}_5
~=~
\frac{1}{2}\left(1- {\delta}_{su}\right)\left(1- {\delta}_{tu}\right) 
 \Biggl\{{stu\choose st0}_5  \left(Y_{a5}-Y_{55} \right)
\nl && 
+~{0st\choose 0st}_5\left({\delta}_{au}-{\delta}_{5u} \right)
-{0st\choose 0su}_5\left({\delta}_{at}-{\delta}_{5t} \right)
-{0ts\choose 0tu}_5\left({\delta}_{as}-{\delta}_{5s} \right) 
\Biggr\} ,
\\ 
\label{eq-wa-88}
{\Sigma}^{3,stu}_{a} 
&\equiv&
\sum_{i=1}^{4}(q_a \cdot q_i)  {0stu\choose istu}_5
~=~
-\frac{1}{2}
\Biggl\{{stu\choose stu}_5  \left(Y_{a5}-Y_{55} \right)
\nl
&& +~{stu\choose 0tu}_5\left({\delta}_{as}-{\delta}_{5s} \right)
+{tsu\choose 0su}_5\left({\delta}_{at}-{\delta}_{5t} \right) 
 +{ust\choose 0st}_5\left({\delta}_{au}-{\delta}_{5u} \right) 
\Biggr\} .
\eea
\newpage
Sums for $1$-point functions:
\bea
\label{eq-wa-89}
{\Sigma}^{1,stuv}_{a} &\equiv&
\sum_{i=1}^{4}(q_a \cdot q_i)  {vstu\choose istu}_5
~=~\frac{1}{2}\left(1- {\delta}_{sv}\right)\left(1- {\delta}_{tv}\right)\left(1- {\delta}_{uv}\right) 
\Biggl\{{stu\choose stu}_5  \left( {\delta}_{av}-{\delta}_{5v}\right)
\nl&&
-~{stu\choose stv}_5\left({\delta}_{au}-{\delta}_{5u} \right)
- {sut\choose suv}_5\left({\delta}_{at}-{\delta}_{5t} \right)
- {tus\choose tuv}_5\left({\delta}_{as}-{\delta}_{5s} \right)
\Biggr\} , 
\\
{\Sigma}^{2,stuv}_{a} &\equiv&
\sum_{i=1}^{4}(q_a \cdot q_i)  {0stuv\choose istuv}_5
~=~-\frac{1}{2}~
\Biggl\{
{stuv\choose stuv}_5 \left(Y_{a5}-Y_{55} \right)+
                       {stuv\choose 0tuv}_5 \left({\delta}_{as}-{\delta}_{5s} \right)
\nl&& +~               {tsuv\choose 0suv}_5 \left({\delta}_{at}-{\delta}_{5t} \right)+
                       {ustv\choose 0stv}_5 \left({\delta}_{au}-{\delta}_{5u} \right)+
                       {vstu\choose 0stu}_5 \left({\delta}_{av}-{\delta}_{5v} \right)  
\Biggr\} .
\label{eq-wa-99}
\eea

The sums (\ref{eq-wa-84a}) - (\ref{eq-wa-99}) vanish whenever two of the indices $s,t,u,v$
are equal. Nevertheless, in order to underline this property, we have occasionally introduced factors $\left(1- {\delta}_{st}\right) \cdots$
in front of the curly brackets when the vanishing of the right hand side of these equations
for equal indices is not so obvious and comes about due to a cancellation. Keeping this in mind
we can give some simpler representations for (\ref{eq-wa-88}) - (\ref{eq-wa-99}) due to the
many scratches in the signed minors. With $w=10-s-t-u$ and $x=15-s-t-u-v=w+(5-v)$, a detailed
investigation shows that
\bea\label{point2}
{\Sigma}^{3,stu}_{a}&&~=~(Y_{w5}-Y_{55})~(q_a \cdot q_w),~~~~~~s,t,u~~~=1, \dots, 4, \\
\label{point1a}
{\Sigma}^{1,stuv}_{a}&&~=~-(q_a \cdot q_v),~~~~~~~~~~~~~~~~~~~~~s,t,u,v=1, \dots, 4, \nn \\
&&~=~~~~~(q_a \cdot q_x),~~~~~~~~~~~~~~~~~~~~~~ v~~~~~~~~~~=5,\nn \\
&&~=~-(q_a \cdot q_v)+(q_a \cdot q_x), ~~~~~s,t,u~~~~=5, \\
\label{point1b}
{\Sigma}^{2,stuv}_{a}&& ~=~0,~~~~~~~~~~~~~~~~~~~~~~~~~~~~~~~~~~~~s,t,u,v~=1, \dots, 4, \nn \\
&&~=~-(q_a \cdot q_x), ~~~~~~~~~~~~~~~~~~~~~s,t,u,v~=5 .
\eea 
Coefficients (\ref{point2}) multiply $I_2(m_w,m_5)$ ($w=1,\dots,4$) and  (\ref{point1a}),(\ref{point1b})
multiply  $I_1(m_5)$ ($s,t,u,v=1,\dots,4$) and $I_1(m_x)$ ($x=1,\dots,4$) if one of the
indices $s,t,u,v$ is equal to $5$.
In the other cases one better keeps the notation in terms of signed minors. 

\vspace{0.5cm}

The above sums are complete in the sense that no more sums occur if the integrals are
contracted with external momenta. Other sums, however, can occur e.g. if the integration momentum is
self-contracted or if, in special investigations, in the above double sums one of the indices remains uncontracted.
Since not all these sums can be dealt with in this letter, we scetch their formal derivation.

In principle, the only relation
needed is found in \cite{Melrose:1965kb},
\bea
{\left( \right)}_n {ik\choose jl}_n={i\choose j}_n {k\choose l}_n-{i\choose l}_n {j\choose k}_n.
\label{master}
\eea
Let us prove \eqref{eq-wa-84a} as an example.
We write
\bea
{\left( \right)}_5 \sum_{i=1}^{4}(q_a \cdot q_i)  {ts\choose is}_5=&&\sum_{i=1}^{4}(q_a \cdot q_i)  
\left[{t\choose i}_5{s\choose s}_5-{s\choose i}_5{s\choose t}_5\right] \nn \\
=&&{s\choose s}_5\sum_{i=1}^{4}(q_a \cdot q_i){t\choose i}_5-{s\choose t}_5 \sum_{i=1}^{4}(q_a \cdot q_i){s\choose i}_5.
\eea
With \eqref{Scalar1} and \eqref{Scalar2} we see that ${\left( \right)}_5$ cancels and \eqref{eq-wa-84a}
is obtained. The factor $\left(1- {\delta}_{st}\right)$ only stresses the fact that for $s=t$
the signed minor ${ts\choose is}_5$ vanishes and so does the sum.

The same procedure also applies for the other sums. Let us look at \eqref{eq-wa-84d}.We have to
take into account that \eqref{master} applies for any $n$, i.e. it is also valid if any row and column
with the same index, say $s$, is scratched. This would give
\bea
{s\choose s}_5 {0st\choose ist}_5={0s\choose is}_5 {st\choose st}_5-{ts\choose 0s}_5 {ts\choose is}_5.
\eea
Such relations are called \emph{extensionals} in \cite{Melrose:1965kb}. We now write
correspondingly
\bea
{s\choose s}_5 \sum_{i=1}^{4}(q_a \cdot q_i)  {0st\choose ist}_5=&&\sum_{i=1}^{4}(q_a \cdot q_i) 
\left[{0s\choose is}_5 {st\choose st}_5-{ts\choose 0s}_5{ts\choose is}_5\right] \nn \\
=&&{st\choose st}_5 \sum_{i=1}^{4}(q_a \cdot q_i){0s\choose is}_5 -{ts\choose 0s}_5
\sum_{i=1}^{4}(q_a \cdot q_i)  {ts\choose is}_5.
\eea
The sums appearing here are \eqref{eq-wa-83b} and \eqref{eq-wa-84a}. Inserting these sums,
some algebra shows that the factor ${s\choose s}_5$ can be canceled and \eqref{eq-wa-84d}
is obtained.

The approach is quite general: we multiply the sums under consideration with the proper
Gram determinant such that an \emph{extensional} of \eqref{master} can be applied. 
This reduces the entries in the signed minors to be summed over such that
the obtained sums have signed minors with less entries and are known from former steps.
The Gram determinant multiplying the original sum must cancel at the end after some algebra.
In this way any sum can be obtained by iteration.

\vspace{0.5cm}

We can now scetch how self-contracted integration momenta can be dealt with. Some ''start-up``
sums are (I.7.16)-(I.7.17),~(I.7.20)-(I.7.22). These sums 
present the type of self-contracted integration momenta. 
The (7.16) and (7.17), e.g., read
\bea
\label{eq-wa-83x}
\sum_{i,j=1}^{4} (q_i \cdot q_j) {0s\choose is}_5{0s\choose js}_5 &&=\frac{1}{2} {s\choose s}_5 \left[
{0s\choose 0s}_5+Y_{55}{s\choose s}_5+2 {s\choose 0}_5{\delta}_{5s}\right],
 \\\label{eq-wa-83y}
\sum_{i,j=1}^{4} (q_i \cdot q_j) {is\choose js}_5 &&=\frac{3}{2} {s\choose s}_5.
\eea
In fact , due to \eqref{fulld4} the sum \eqref{eq-wa-83x} is already one of the sums
occurring if the vectors $q_i$ and $q_j$ are contracted. A further sum might be
\bea
\label{eq-wa-83z}
\sum_{i,j=1}^{4} (q_i \cdot q_j) {0si\choose 0sj}_5.
\eea
With
\bea
{s\choose s}_5 {0si\choose 0sj}_5={0s\choose 0s}_5 {is\choose js}_5-{0s\choose is}_5 {0s\choose js}_5
\eea
we see that \eqref{eq-wa-83z} can be reduced to \eqref{eq-wa-83x} and \eqref{eq-wa-83y} with the final result
\bea
\label{eq-wa-83zfin}
\sum_{i,j=1}^{4} (q_i \cdot q_j) {0si\choose 0sj}_5={0s\choose 0s}_5-\frac{1}{2}{s\choose s}_5 Y_{55}-
{s\choose 0}_5 \delta_{5s}.
\eea
In this manner the self-contracted integration momenta can be dealt with like 
the other ones, coming from contractions with external momenta, and thus provide
a consistent picture of our approach.


\end{document}